\documentclass[twocolumn]{aastex631}

\usepackage{mathptmx}
\usepackage[T1]{fontenc}
\usepackage{ae,aecompl}
\usepackage{graphicx}	
\usepackage{amsmath}	
\usepackage{amssymb}	
\usepackage{subfigmat}
\usepackage{enumerate}
\usepackage{multirow}
\usepackage{natbib}

\def\gtrsim{\mathrel{\hbox{\rlap{\hbox{\lower5pt\hbox{$\sim$}}}\hbox{$>$}}}}

\newcommand{\methanol}{\mbox{CH$_3$OH}}
\newcommand{\degree}{\mbox{$^\circ$}}
\newcommand{\rom}[1]{\uppercase\expandafter{\romannumeral#1}}

\newcolumntype{Y}{>{\centering\arraybackslash}X}

\shorttitle{Variability of COMs Emission in B335}
\shortauthors{Lee et al.}

\begin{document}

\title{A Natural Laboratory for Astrochemistry, a Variable Protostar B335}

\author[0000-0003-3119-2087]{Jeong-Eun Lee}
\affil{Department of Physics and Astronomy, Seoul National University, 1 Gwanak-ro, Gwanak-gu, Seoul 08826, Korea}
\affil{SNU Astronomy Research Center, Seoul National University, 1 Gwanak-ro, Gwanak-gu, Seoul 08826, Republic of Korea}
\email{lee.jeongeun@snu.ac.kr}

\author{Neal J. Evans II}
\affiliation{Department of Astronomy, The University of Texas at Austin, 2515 Speedway, Austin, TX 78712, USA}

\author{Giseon Baek}
\affil{Department of Physics and Astronomy, Seoul National University, 1 Gwanak-ro, Gwanak-gu, Seoul 08826, Korea}

\author{Chul-Hwan Kim}
\affil{Department of Physics and Astronomy, Seoul National University, 1 Gwanak-ro, Gwanak-gu, Seoul 08826, Korea}

\author{Jinyoung Noh}
\affil{Department of Physics and Astronomy, Seoul National University, 1 Gwanak-ro, Gwanak-gu, Seoul 08826, Korea}

\author[0000-0001-8227-2816]{Yao-Lun Yang}
\affiliation{Star and Planet Formation Laboratory, RIKEN Cluster for Pioneering Research, Wako-shi, Saitama, 351-0106, Japan}

\begin{abstract}

Emission lines from complex organic molecules in B335 were observed in four epochs, spanning a luminosity burst of about 10 years duration. The emission lines increased dramatically in intensity as the luminosity increased, but they have decreased only slightly as the luminosity has decreased. This behavior agrees with expectations of rapid sublimation as the dust temperature increases, but slower freeze-out after the dust temperature drops. Further monitoring of this source, along with detailed chemical models, will exploit this natural laboratory for astrochemistry.
\end{abstract}


\section{Introduction}

Complex Organic Molecules (COMs), defined as carbon-bearing species with 6 or more atoms \citep{herbst2009, Ceccarelli2023PPVII}, are the building blocks of prebiotic molecules, such as amino acids and sugars, essential for terrestrial life. The recent Rosetta mission showed that many COMs, as well as water and prebiotic molecules, exist in the comet 67P/C-G \citep{Altwegg2017}, revealing that the early solar nebula was rich in water and organic molecules. The comets share abundance patterns with COMs seen in heated regions  in protostellar envelopes and disks around forming stars \citep{Drozdovskaya2019, jelee19, Jorgensen2020, 2019ESC.....3.2659B, 2020EPSC...14..628P, 2021NatAs...5..684B, 2024ApJ...970L...5L}.
Thus, studying COMs in protostars provides a link to the initial ice compositions of comets.

Recent JWST observations of low mass protostars, the precursors of Sun-like stars, reveal a rich and complex ice inventory.
Most ice features of COMs are mixed with various ice components, which makes it challenging to decompose individual components robustly \citep{Yang2022,Rocha2024, Chen2024}. In addition, the ice features can only trace the ice composition outside their sublimation radius. 
Radio wavelength observations have identified an even richer collection of COMs on small scales where the dust temperature has exceeded the sublimation temperature of water (100 K). 
These objects are hot ($>$ 100 K), dense ($>$ $10^7$ cm$^{-3}$), and compact ($<$ 100 au) and identified as hot corinos \citep{Ceccarelli2004, Ceccarelli2007}.
While single-dish surveys have provided an inventory 
(e.g., \citealt{2003ApJ...593L..51C, 2004ApJ...615..354B}),
the high-resolution observations with ALMA
have shown the extent of the hot corinos \citep[e.g.,][]{Jacobsen19, Sahu2019, shlee20, Okoda2022, jelee2023, jelee2024}, which are not always consistent with the current luminosity of the protostars.
The cases with hot corinos more extended than the water sublimation radius at the current luminosity
(e.g., \citealt{vanthoff2022})
along with evidence for CO sublimation in very low luminosity protostars 
\citep{2012ApJ...758...38K},
suggested higher luminosities in the past, as was predicted by models of episodic accretion leading to luminosity bursts \citep{Dunham12}. The chemical response to these luminosity bursts has been modeled \citep{jelee2007}.
If we could observe such a burst, it would provide a natural laboratory to study the processes of ice sublimation and subsequent evolution.

Over eight years of comprehensive all-sky monitoring, NEOWISE (Near-Earth Object Wide Infrared Survey Explorer) has enabled meticulous examination of mid-infrared light curves from thousands of protostars within the Gould Belt \citep{wspark2021, slee2024}. Approximately 20\% of the analyzed Class 0/I sources exhibit prolonged fluctuations in brightness over time, indicating substantial evidence of accretion variability. For those sources monitored in both mid-infrared and sub-millimeter wavelengths, observed brightness changes are closely correlated \citep{carlos2020, yhlee2020}, validating the expectation that significant mid-infrared variations track  changes in luminosity and, consequently, accretion rates \citep{Dunham12}. 

One of the variable Class 0 sources identified with WISE/NEOWISE, B335, experienced a luminosity increase of a factor of 5 to 7 over the timescale of $\sim$10 years \citep{evans2023, chkim2024}. We had a series of ALMA observations with similar spectral coverage serendipitously spaced during the ascending phase of luminosity. We also intentionally performed an observation to cover a similar spectral range to the previous observations in the descending phase. Spectra have been obtained with JWST, both NIRSpec and MIRI, also during the descending phase 
(\citealt{2024ApJ...966...41F}, Rubinstein et al., in press). 
Therefore, the chemical behavior during the luminosity burst is exceptionally well documented.  B335 is a well-studied Class 0 object with $T_{\rm bol} = 46$ K \citep{Yang2018} which has been modeled in detail \citep{evans2023}, providing dust temperature estimates as a function of radius and luminosity.
These features make B335 a (so far) unique laboratory in which to study the natural experiment of ice sublimation and COMs release in real time.

\section{B335: a Very Young variable protostar}

The Bok globule B335 is a well-studied, isolated dense core with a deeply embedded Class 0 protostar. A scattered light region around the core links it to a nearby A2 star, HD184982,
with a Gaia DR3 distance of $164.5\pm 1.2$ pc, an unusually well-known distance for an isolated globule \citep{Watson2020}. It was the first source with infall detected from line profiles \citep{Zhou1993} and ALMA observations showed red-shifted absorption against the continuum, definitive evidence for infalling gas \citep{evans2015}. Rotation is slow ($\Omega < 4 \times 10^{-14}$ s$^{-1}$), and any Keplerian disk must be less than about 16 au in size \citep[][after correction to the new distance]{Yen2015}, making a good case for magnetic braking. 
Complete spectral scans of the source were made with Herschel PACS and SPIRE \citep{Yang2018}, and maps at submillimeter wavelengths constrain the radial density distribution of the envelope \citep{shirley2002}. Models of the source, including an infalling envelope, a disk, and outflow cavities, produce a good match of the full SED and radial intensity distributions for an inside-out collapse with age since initiation of collapse of $(4 \pm 1) \times 10^4 $ years \citep{shirley2002, evans2023}, consistent with the outflow age \citep{Hirano1988} corrected to the new distance. Using the complete Spitzer and Herschel SEDs and the inclination angle of 87 degrees \citep{Stutz2008}, the pre-burst central luminosity of 3 L$_{\odot}$ was derived \citep{evans2023}. This luminosity predicts a water snow line at $\sim$25 au \citep{Bisschop2007,vanthoff2022} in the envelope; a very compact region ($r< 15$ au) of emission of COMs was found by \citet{Okoda2022}. 

Then, something happened. Between MJD 55304 (April 2010) and MJD 56948 (October 2014), the source brightened substantially in the near-IR, reaching the maximum in September 2018 (around MJD 58400). This increase was recognized only recently by collecting WISE and NEOWISE data. Increasing the luminosity in the pre-outburst model by a factor of 5-7 reproduces the WISE (W2) light curve \citep{evans2023, chkim2024}.

\begin{figure}[tbh]
\centering
\includegraphics[height=4.5in]{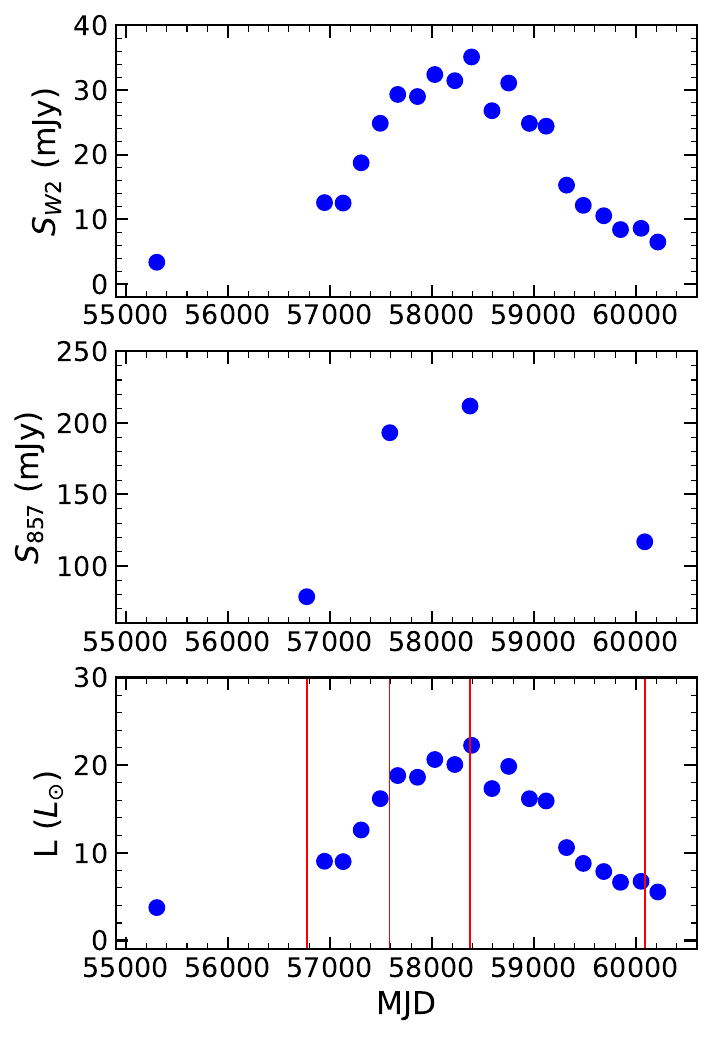}
\caption{Light curves of B335. \textit{Upper : }The mid-IR flux variation of B335 observed with WISE/NEOWISE in W2 (4.6 $\mu$m) of B335.
\textit{Middle : }Four epochs fluxes at 857 $\mu$m observed by ALMA. Fluxes were measured over the aperture of 2\arcsec.
\textit{Bottom : }The variation of luminosity of B335 obtained using the variation of mid-IR flux density. The relation between luminosity and mid-IR flux density was explored by \citet{evans2023}. The vertical red lines indicate the dates of Band 7 ALMA observations. }
\label{fig:lightcurve}
\end{figure}

\section{Observations}
\subsection{WISE/NEOWISE}
B335 was observed with the WISE survey in April 2010 and has been observed by the NEOWISE survey every six months since 2014. The WISE survey of B335 consists of four bands: W1 (3.4 \micron), W2 (4.6 \micron), W3 (12 \micron), and W4 (22 \micron), while the NEOWISE survey of B335 consists of two bands: W1 (3.4 \micron) and W2 (4.6 \micron). NEOWISE survey has been conducted 10-20 observations in an epoch over 1-3 days. We explored the NASA/IPAC Infrared Science Archive to construct the light curve of B335 using a query radius of 5\arcsec\ around the coordinates of B335. We adopted the averaging method from \citet{park21} to average the multiple observations at one epoch into a single data. 
Figure \ref{fig:lightcurve} (top) shows the derived light curve of W2 in flux. 

\subsection{Previous ALMA observations}
In order to investigate the variability of the COMs emission in response to the changing luminosity, we retrieved the ALMA band 7 data of B335 (2012.1.00346.S and 2015.1.00169.S PI: Neal J. Evans \rom{2}) from the ALMA archive. The 2012.1.0034.6.S observation was performed on 2014 April 27, while the 2015.1.00169.S observations were carried out in 2016, July 17 -- 23, and in 2018, September 11 and 18. The 2012.1.00346.S observation has four spectral windows (SPWs), and each SPW has a channel width of 0.061 MHz with a bandwidth of 234.375 MHz. The spectral resolution is two times the channel width. Refer to \citet{evans2015} for the detailed spectral setup for the 2012.1.00346.S observation. The 2015.1.00169.S observation has four SPWs;
each SPW contains the HCN 4-3, HCO$^{+}$ 4-3, CS 7-6, and H$^{13}$CN 4-3 with a channel width of 0.122 MHz and a bandwidth of 234.375 MHz, except the SPW including H$^{13}$CN with a channel width of 0.488 MHz and a bandwidth of 468.750 MHz.

\subsection{New ALMA observations}
\begin{figure*}[htp]
    \centering
    \includegraphics[width=1\textwidth]{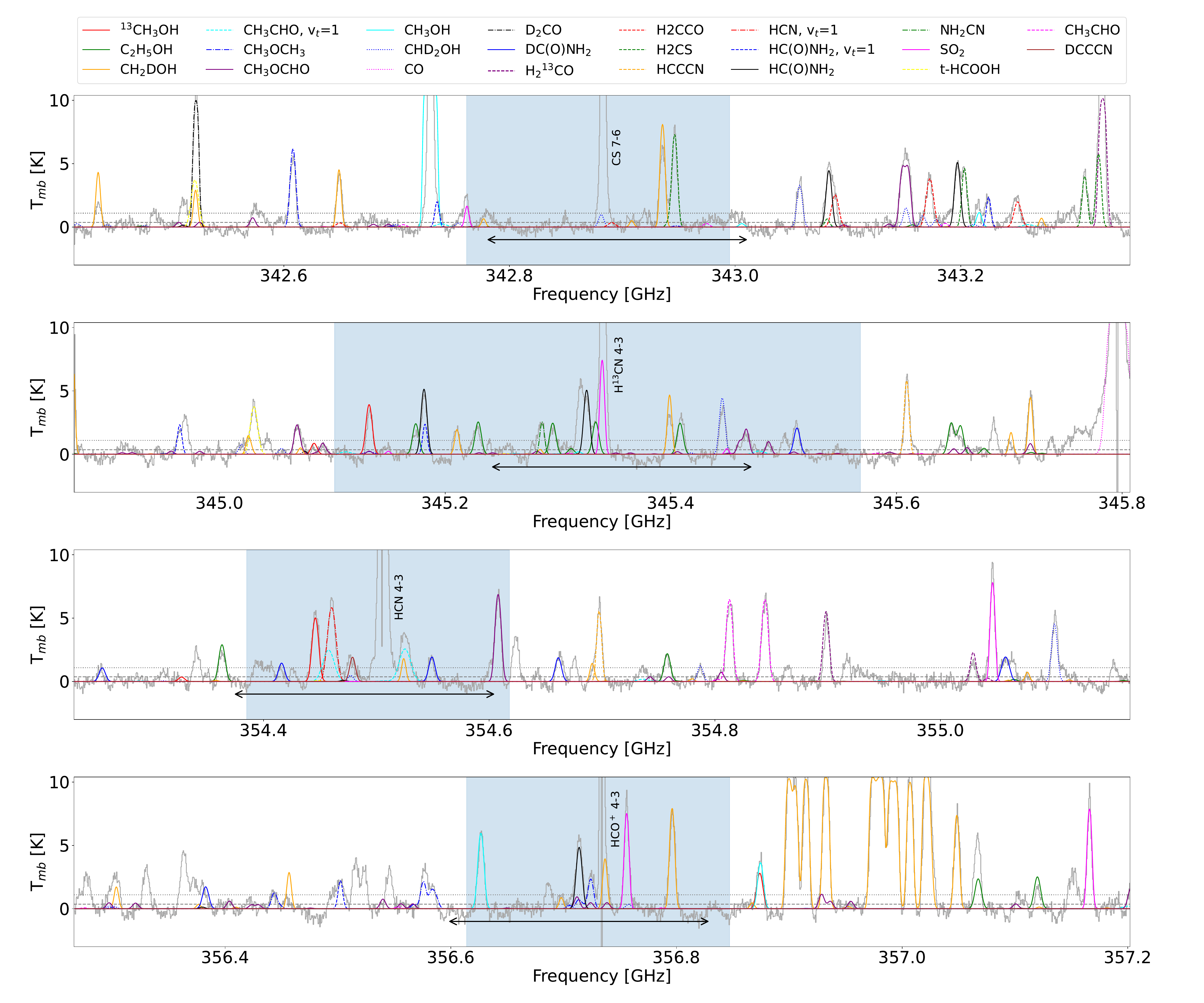}
    \caption{The spectra of B335 for the 2023 epoch. 
    The gray spectra represent the observed spectra, while the colored lines show the fitted spectra using XCLASS. Different colors indicate different species, as marked at the top. All identified lines from the XCLASS fitting above the 3$\sigma$ threshold are listed in Table \ref{tab:Identified_Lines} of the Appendix. More than 10 lines remain unidentified.
    The shaded region indicates where the 2016, 2018, and 2023 data were fitted to obtain the abundance and column density presented in Figure \ref{fig:abundance}. The horizontal black arrows denote the spectral range of the 2014 data, which is significantly narrower than that of the other epochs. Various molecules are distinguished by different colors and line styles. 
    The dashed and dotted grey horizontal lines represent the 1$\sigma$ and 3$\sigma$ levels, respectively.}
    \label{fig:all_range}
\end{figure*}

A new observation of B335 was performed during Cycle 9 (2022.1.00986.S; PI: Giseon Baek) on 2023 May 25 and included two observations. The coordinates of the phase center for two observations are $\alpha=19^{h}37^{m}00^{s}.894$ and $\delta=+07^{d}34^{m}09^{s}.590$ (J2000). The observation in Band 7 was carried out with a baseline range from 27.1 m to 3.6 km, with 42 antennas. The spectral windows (SPWs) were set to cover the frequency range of previous observations (2012.1.00346.S and 2015.1.00169.S) to investigate the variability of the COMs emission with luminosity change, although the spectral coverage is wider than those of previous observations (Figure \ref{fig:all_range}). Like previous observational data, the SPWs contain CS 7-6, H$^{13}$CN 4-3, HCN 4-3, and HCO$^{+}$ 4-3, and all SPWs have the same setup with a bandwidth of 937.500 MHz and a channel width of 0.488 MHz, which is larger than that of the first epoch data by a factor of 8.

The full spectral coverage of the new data is presented in Figure \ref{fig:all_range}, together with the fitted individual spectra. 
We note that the first epoch's spectral coverage is narrower than those of the other three epochs, especially around the H$^{13}$CN 4-3 line. 
The shaded regions in Figure \ref{fig:all_range} present the spectral coverage of the second and third epochs, while the horizontal arrows indicate the spectral coverage of the first epoch.
  
\section{Data reductions}
To achieve better imaging, self-calibration was applied for all the data using the CASA 6.2.1 pipeline \citep{mcmullin2007}. The 2022.1.00986.S was calibrated using CASA 6.4.1 pipeline \citep{mcmullin2007} instead. The 2022.1.00986.S was also self-calibrated to obtain better imaging using the same CASA version. 
\par
The UV distance for 2012.1.00346.S is shorter than the UV distances for 2015.1.00169.S and 2022.1.00986.S. To directly investigate the variability of the COMs emission along the variable luminosity, we used the same parameters, such as \texttt{UVrange} and restoring beam, when we performed the \textit{tclean} task. In the CLEAN process, we adopted the Hogbom algorithm with a robust parameter of 0.5 for all the data. The adopted UVrange parameter is 22.2 to 558.2 m, corresponding to the UV distance for 2012.1.00346.S, and the adopted restoring beam parameter is 0\farcs45 $\times$ 0\farcs40 with a position angle of 78$\degree$. 
Finally, we binned channels to match the spectral resolution to the last epoch data.

\begin{figure}[tbh]
\centering
\includegraphics[width=0.5\textwidth]{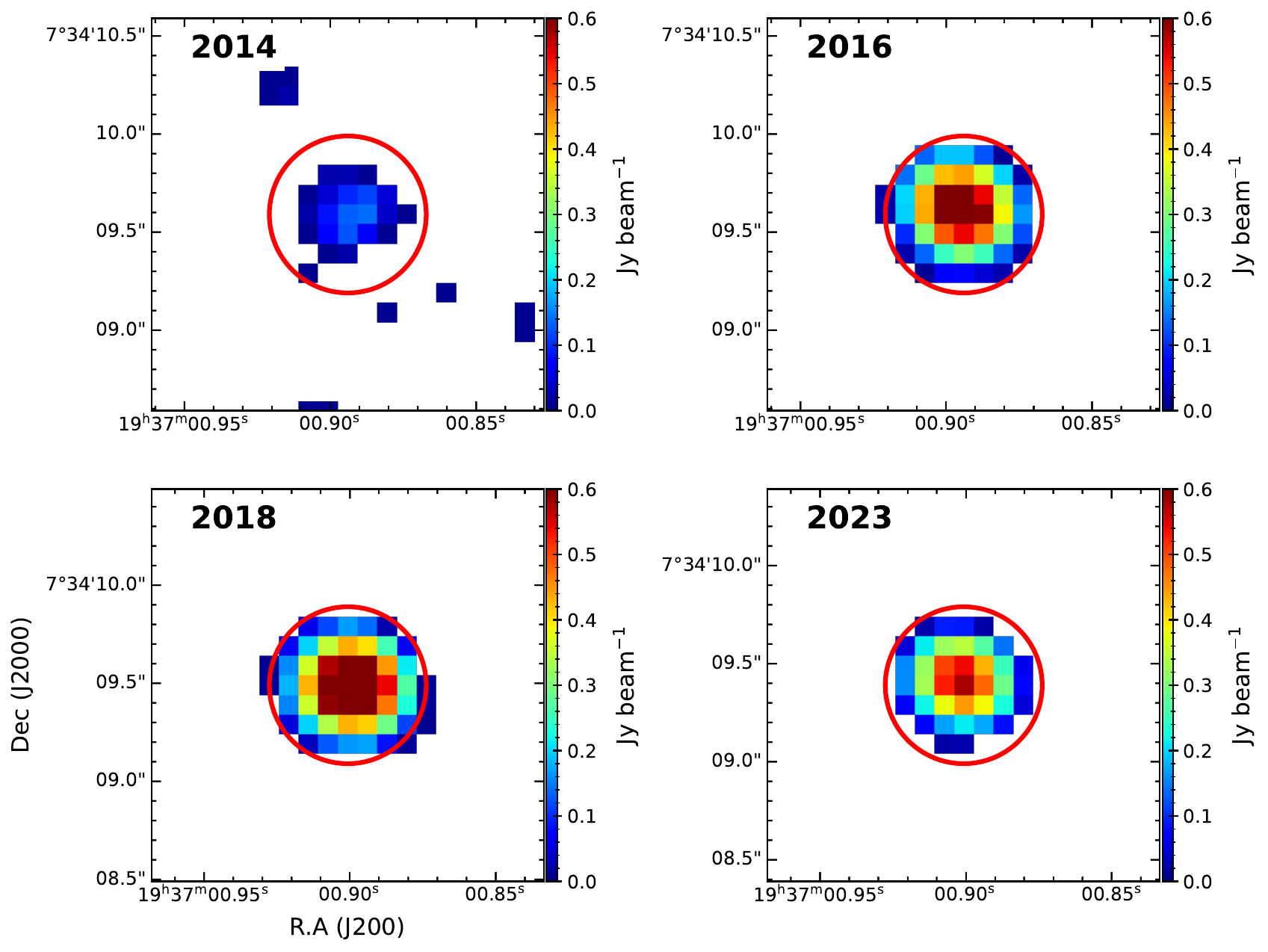}
\caption{Moment 0 map of the \methanol\ {\bf $23(4,20) - 22(5,18)$, E v$_t$=0} 
line in four epochs. Each moment 0 map uses only emission above 3$\sigma$ (1$\sigma$ = 8.6 mJy beam$^{-1}$ in Epoch 1 was adopted).
The red circles in the moment map have the same diameter of 0.8\arcsec. The synthesized beam for each data set is located in a gray oval in the lower left corner.}
\label{fig:ch3oh_mom0}
\end{figure}

In summary, after the self-calibration, we performed the \textit{tclean} task with the same parameters to image all four data sets to have the same spatial resolution of 0\farcs45 $\times$ 0\farcs40. After binning, the final spectral resolution is 0.4 km s$^{-1}$, which is much smaller than the typical COMs line width of $\sim$5 km s$^{-1}$, in all four data sets.
Figure \ref{fig:ch3oh_mom0} presents the integrated intensity maps of a CH$_3$OH line in all four epochs to show a dramatic increase in methanol emission between 2014 and 2016.
The spectra extracted from the continuum peak for all four epochs are compared in Figure \ref{fig:spectra}. The rms noises of the spectra in all epochs are similar, ranging from 7 mJy beam$^{-1}$ to 10 mJy beam$^{-1}$.

The continuum fluxes for the four epochs are measured with an aperture size of 2\arcsec\  centered on the peak position.
The continuum emission at 345 GHz shows a dramatic increase (Figure \ref{fig:lightcurve}, middle) during the first three epochs but drops at the fourth epoch significantly, which is consistent with the NEOWISE light curves.
The fluxes of all molecular emissions also increase with time during the first three epochs, but the most interesting development is the dramatic increase in the COMs lines (Figure \ref{fig:spectra}).

\section{The Variability of COMs Emission}

\begin{figure}
\centering
\includegraphics[width=0.47\textwidth]{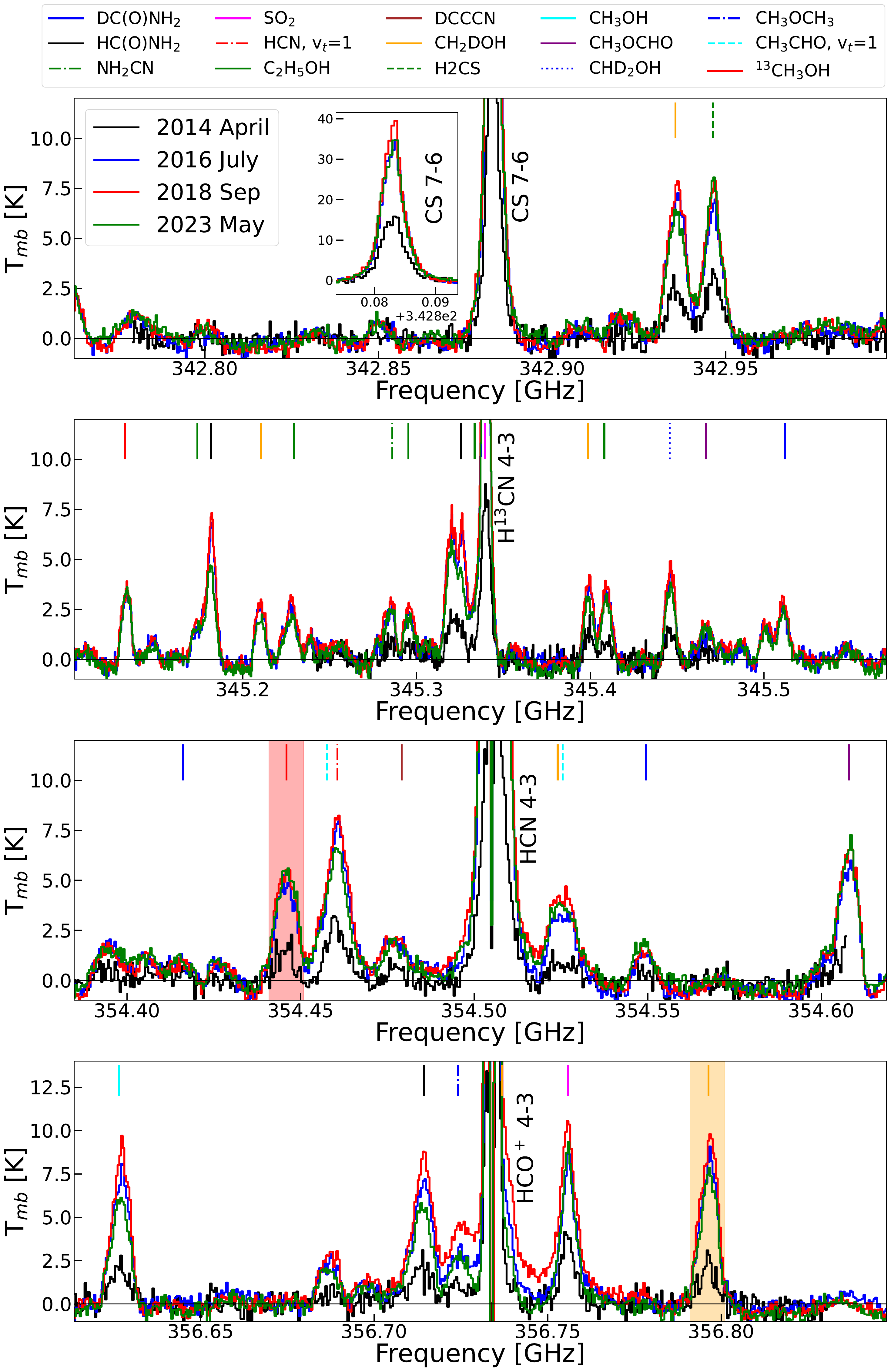}\\
\caption{Spectra of B335 observed in four epochs. Black, blue, red, and green solid lines depict the observed spectra in each epoch. Vertical lines indicate the individual COMs modeled by XCLASS. The inset at the top panel shows the overall line profiles of CS J$=7-6$ (E$_u$=65.8 K). The red- and orange-shaded spectra are 
$^{13}$CH$_3$OH (E$_u$=43.7 K) and CH$_2$DOH (E$_u$=234 K) lines, respectively, with very different upper state energies.}
\label{fig:spectra}
\end{figure}

The ALMA band 7 observations in three epochs through the brightening phase shows {\em the first detection of COMs variability in response to the protostellar luminosity change (Figures \ref{fig:ch3oh_mom0} and \ref{fig:spectra})}. 
The COMs intensities obtained from the continuum peak position
(Figure \ref{fig:spectra}) increase as the mid-IR and submillimeter fluxes increase (Figure \ref{fig:lightcurve}) during the first three epochs (black, blue, and red spectra for 2014, 2016, and 2018, respectively). On the other hand, the COMs intensities (green spectra) at the fourth epoch (2023), when the protostellar luminosity decreased to the level of the first epoch (2014), did not decrease proportionally to the continuum flux, although they did show a slight decrease of about 30~\%, for example, in the \methanol\ line, the leftmost isolated line at 356.627 GHz in the bottom panel of Figure \ref{fig:spectra}.
This apparent flux variability of COMs in B335 allows us to investigate the relative variation in different species.

To investigate the variability of COM abundances, we first derived the gas temperature variation caused by the luminosity variation using the excitation diagrams of detected COMs.
The derived temperatures were adopted to fit the observed spectra using XCLASS \citep{moller2017} under the LTE assumption. The line width was calculated from the observed spectra.

The limited number of transitions makes it difficult to estimate accurate excitation temperatures of all detected COMs.
Since all the COMs detected in the first epoch of 2014, except for CH$_2$DOH, had fewer than three transitions in our spectral range, we adopted the excitation temperature of CH$_2$DOH in all four epochs for consistency. 
According to our best-fit XCLASS results, the optical depths of all COM transitions presented in Figure \ref{fig:spectra} are less than 1, supporting the adequacy of our excitation diagram analysis.
The CH$_2$DOH excitation temperatures in the four epochs are presented in Figure \ref{fig:abundance} (lower panel) and Table \ref{table:Ncom} in the Appendix.
According to the higher resolution images by \citet{Okoda2022}, however, all COMs do not trace the same region; in particular, nitrogen-bearing COMs trace more inner gas than other oxygen-bearing COMs.  Therefore, the column densities of HC(O)NH$_2$ and DC(O)NH$_2$ derived with the CH$_2$DOH excitation temperature are likely underestimated.

The raised gas temperature alone can also increase the line intensity even with the same number of molecules. This was the case for HCO$^+$, for which the J$=4-3$ transitions from 2012 and 2018 could both be fitted with the same abundance, varying only the luminosity and hence temperature \citep{evans2023}. However, the increase in that transition was only a factor of two. 
We compare the COMs to CS, whose sublimation temperature ($\sim$60 K, \citet{jelee2004, Ferrero2020, Perrero2022}) is much lower than the calculated gas temperature at all four epochs; consequently, all CS likely ends up in the gas phase at the hot central region where the density is high enough to generate the CS J$=7-6$ line. CS is very sensitive to the gas temperature since its ground state energy (E$_u$) is as high as 65.8 K. Although images for its intensity distribution are not presented in this paper, except for the outflow cavity, the emission region of CS J$=7-6$ around the central source is not resolved, like the methanol line (Figure \ref{fig:ch3oh_mom0}), probably because of its high critical density of $2\times 10^7$ cm${-3}$ \citep{plume1922}.
Therefore, the enhanced CS emission can be attributed to only the raised temperature, and the integrated intensity (or the peak intensity) of CS J$=7-6$ increased only by a factor of 2.3 from 2014 to 2018, as presented in the inset at the top panel of Figure \ref{fig:spectra}. In contrast, the integrated intensities of both $^{13}$CH$_3$OH (E$_u$=43.7 K at 356.446 GHz) and CH$_2$DOH (E$_u$=234 K at 356.796 GHz) lines (shaded spectra at Figure \ref{fig:spectra}) increased by a factor of 4.7 from 2014 to 2018.  Even if some CS is added to the gas phase by the luminosity increase, the larger enhancement factor for the COMs lines strongly indicates that the enhanced strength of COM lines over the burst is mainly caused by the expanded sublimation region of COMs, i.e., the increase in the total numbers of COMs in gas. However, we need a model to accurately separate the effects of temperature and column density on the enhanced line strength of COMs.

\begin{figure}
\centering
\includegraphics[width=0.5\textwidth]{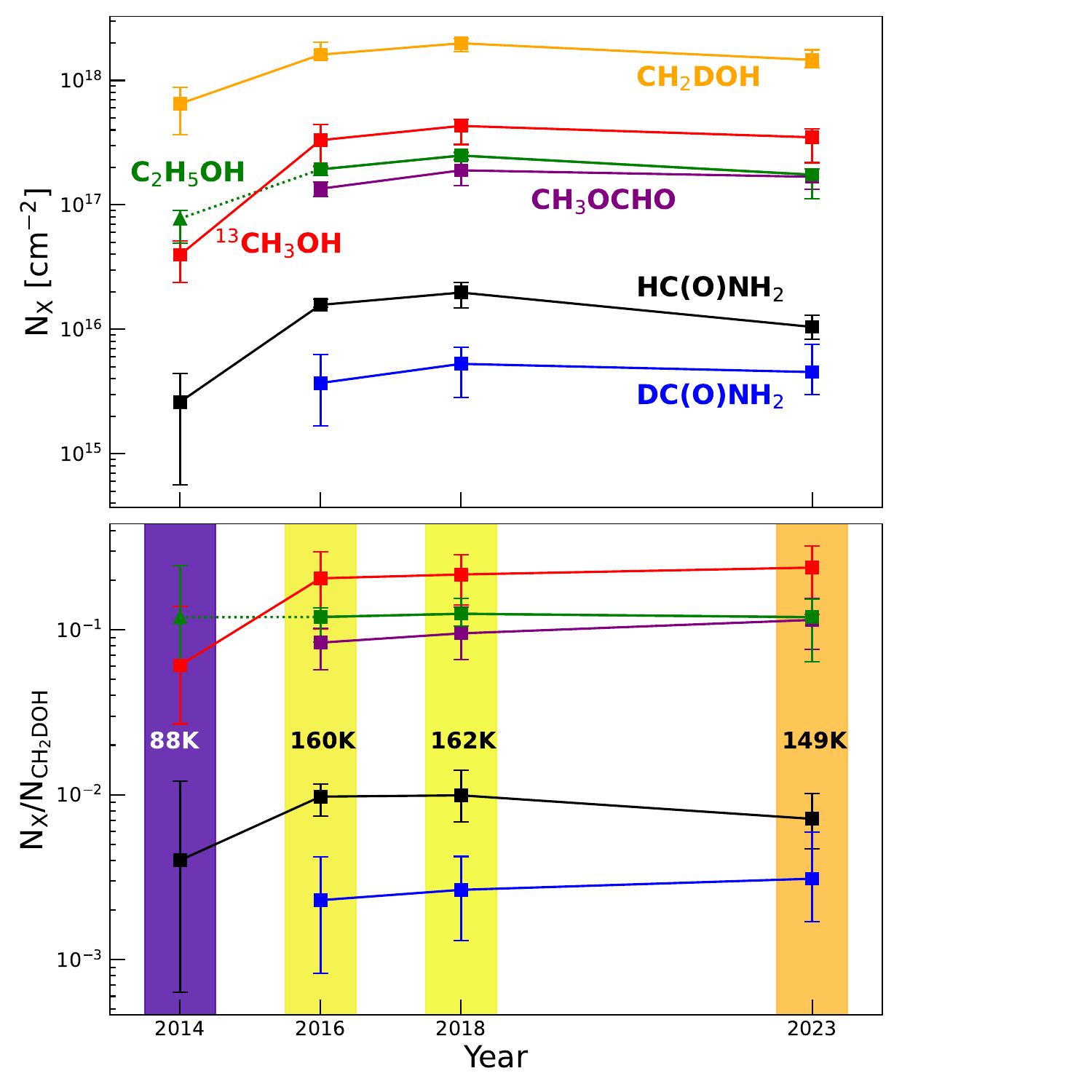}
\caption{Column density (upper panel) and abundance (lower panel) variability of COMs derived from the XCLASS fitting of four epochs data in 2014, 2016, 2018, and 2023. The abundance of each species was calculated relative to CH$_2$DOH. 
The numbers within the color bars present the excitation temperatures of CH$_2$DOH in four epochs.}
\label{fig:abundance}
\end{figure}

We obtained the column density variation of each molecule by fitting the observed spectra using XCLASS. 
The source size ($FWHM=0.12\arcsec$) during the burst was estimated from the 2D Gaussian fit of the CH$_3$(OH)CHO and HC$_3$N high-resolution images obtained in 2019 by the ALMA program: 2018.1.01311.S \citep{Okoda2022, okoda2024}. We adopted the same source size for the pre-burst data obtained in 2014, which was not spatially resolved. The derived column density might be underestimated if the emission size was significantly smaller than 0.12\arcsec. However, the column density ratio between species, i.e., abundance, will eliminate the effect of the emission size. In Figure \ref{fig:abundance} (upper panel), the different color lines show the column density variation of different COMs, with two or more lines detected, during all four epochs. The column densities of all COMs detected in B335 are listed in Table \ref{table:Ncom} in the Appendix.
The plot at the lower panel of Figure \ref{fig:abundance} shows the corresponding variation of abundances relative to CH$_2$DOH, which includes the largest number of lines detected in 2014. Note that only one CH$_3$OH line was covered by the spectral range of the first epoch. Thus, the derived column density of CH$_3$OH may not be as accurate as other species.
Although the abundances of two molecules ($^{13}$CH$_3$OH and HC(O)NH$_2$) detected in 2014 seem lower than those in later epochs, they are still within the margin of error. Therefore, the abundances of all COMs, relative to CH$_2$DOH, presented in Figure \ref{fig:abundance} do not show a notable variation over the four epochs, given the uncertainties.

\section{Discussion}

In B335, the deuterated species such as CH$_2$DOH and DC(O)NH$_2$ are firmly detected, and the D/H ratios of CH$_3$OH and HC(O)NH$_2$, derived from the spectra observed in 2023, are greater than 0.1, indicative of very cold phase chemistry. This is consistent with the results by \citet{okoda2024}. In addition, B335 is too young to have lost the prestellar chemical condition \citep{evans2023}.
Another very young protostar with a recent burst event, HOPS 373 SW, also revealed a very high D/H ratio of CH$_3$OH, and thus, is considered a good laboratory for the study of ice composition in prestellar cores \citep{jelee2023}. However, B335 is the first and only low-mass protostar for which we have monitored a real-time change of the COMs line strength throughout a burst event from the pre-burst phase. Thus, it is a unique test bed for studying the sublimation of COMs from dust grains and the freeze-out of COMs back to grain surfaces in episodic accretion in low-mass star formation.

During the brightening phase, the location of snow lines is pushed outwards, and thus the line fluxes increase due to the immediate increase of the total mass of sublimates. 
The size of the water snow line in a 1D spherical envelope increases with luminosity by $R_{\rm snow} \sim R(100~{\rm K}) = 15.4 \sqrt{L/L_{\odot}}$ au \citep{Bisschop2007,vanthoff2022}. This formula predicts a snow line in the protostellar envelope with the maximum luminosity of 22 $L_{\odot}$ in 2018 at $\sim$72 au, versus 40 au in 2014, with the luminosity of 6 $L_{\odot}$.
In contrast, the resolved radius of COMs emission during the burst is about 20 au, much smaller than these predictions. Using 3D models with outflow cavities and an embedded disk \citep{evans2023}, we find a snow line with a maximum extent of $\sim$20 au, even for the peak luminosity of 22 $L_{\odot}$, consistent with the observations. The outflow cavity channels short wavelength radiation away from the COM-rich envelope and the disk blocks radiation in the mid-plane, resulting in a very non-spherical and smaller hot corino.

Even after the burst ceases, the imprints of the burst survive in gas phase chemistry while the dust properties immediately respond to the luminosity changes \citep{jelee2007,Johnstone2013}; the radiative heating timescale of dust is the same as the light traveling time to the snow line ($\sim$20 au), i.e., about 15 hours \citep{jelee2004}. 
Meanwhile, the sublimates are destroyed by gas-phase reactions in several $10^4$ years \citep[e.g.,][]{nomura2009}, which is much longer than burst timescale ($\sim 100$ years for an outburst, \citet{jelee2007}).
Therefore, the derived abundances of COMs relative to CH$_2$DOH preserve the ice composition. The latest burst in B335 occurred over about 10 years from start to end. 
Therefore, a more important chemical process responding to the years-long luminosity variability is the freeze-out of COMs onto grain surfaces. Chemical modeling of outbursting protostars also finds a wide range of retention timescales in organics \citep{Molyarova2018}, which requires constraints from monitoring observations after a burst event ends.

According to our monitoring result, the column densities in 2023 remained the same as in 2016 and 2018, although the luminosity returned to the pre-burst level in 2023. This suggests that the freeze-out of COMs did not occur significantly between 2018 and 2023, even though the luminosity has decreased to the pre-burst level. From the best-fit model for B335, the density at 20 au, the snow line location at the maximum burst phase, is at least 10$^8$ cm$^{-3}$ \citep{evans2023}. Assuming the gas temperature of 150 K, which reproduces the quiescent phase COMs emission well, the freeze-out timescale is about 30 years \citep[see Table 4 in][]{jelee2004}. This timescale is the upper limit assuming pure infalling envelope distribution. If the density is 5$\times$10$^8$ cm$^{-3}$, or a small disk exists, then the COMs will freeze out faster back onto grain surfaces, in $\sim$5 years \citep{jelee2004,vanthoff2022}.
Indeed, the comparison of models to ALMA continuum emission suggests an additional dense structure on the scale of tens of au \citep{evans2023}.
However, we do not see a significant column density decrease in any COM between 2018 and 2023. 
Therefore, further monitoring of B335 is critical to investigate the chemical variability after the end of a burst event; the current luminosity is close to the pre-burst level. In particular, it will provide a unique chance to measure the in-situ freeze-out timescale. 

In addition, when the burst ended in 2023, the excitation temperature did not change much compared to 2016 and 2018. This indicates that the gas cooling timescale is much longer than the dust cooling timescale; the decrease in continuum flux in 2023, corresponding to the change in luminosity, indicates that the dust cooling process is efficient and instantaneous.
The excitation temperature derived from fitting multiple lines will measure the kinetic temperature only if the levels have reached equilibrium with the translational degrees of freedom via collisions. In addition, the kinetic temperature will track the dust temperature only if collisions of gas particles with the dust have brought the translational degrees of freedom into equilibrium with the dust temperatures. If the excitation temperatures remain elevated above the dust temperatures, obtained from detailed modeling, future observations may allow constraints on the timescales associated with these processes.

\section{Conclusion}
We monitored the variability of COM emission in response to changes in protostellar luminosity during a recent burst event in B335, spanning 10 years from the pre-burst phase in 2014 to the end of the burst in 2023.
Our analysis of the monitoring observations reveals that COMs sublimate instantaneously, increasing their column densities in the gas phase in response to the luminosity rise during the burst event.
However, the decrease in COM column densities during the declining phase, up to the end of the burst, is minimal, suggesting a freeze-out timescale longer than 5 years.
Therefore, continued monitoring of COM emission variability in B335 is crucial to accurately measure the in-situ freeze-out timescale of COMs.

\section{Acknowledgements}

This work was supported by the New Faculty Startup Fund from Seoul National University. This work was also supported by the National Research Foundation of Korea (NRF) grant funded by the Korea government (MSIT) (grant numbers 2021R1A2C1011718 and RS-2024-00416859).
GB was supported by Basic Science Research Program through the National Research Foundation of Korea (NRF) funded by the Ministry of Education (RS-2023-00247790).
NJE thanks the Astronomy Department of the University of Texas for research support.
Y.-L.Y. acknowledges support from Grant-in-Aid from the Ministry of Education, Culture, Sports, Science, and Technology of Japan (20H05845, 20H05844, 22K20389), and a pioneering project in RIKEN (Evolution of Matter in the Universe).
This paper makes use of the following ALMA data: ADS/JAO.ALMA\#2012.1.0034.6.S, 015.1.00169.S, and 2022.1.00986.
ALMA is a partnership of ESO (representing its member states), NSF (USA) and NINS (Japan), together with NRC (Canada), NSC and ASIAA (Taiwan), and KASI (Republic of Korea), in cooperation with the Republic of Chile. The Joint ALMA Observatory is operated by ESO, AUI/NRAO and NAOJ.

\bibliographystyle{aasjournal}
\bibliography{ms}{}


\appendix

The column densities derived from the XCLASS fitting are summarized in Table \ref{table:Ncom}. 
Note that the spectral coverage of the second (2016) and third (2018) epochs (the shaded region in Figure \ref{fig:all_range} was used to compare column densities over four epochs.
Although some species are identified in Figure \ref{fig:all_range}, the column densities of the species are not listed in this table if all of their detected lines are overlapped with other molecular lines. 
We also list all lines identified from the spectra observed in 2023 in Table \ref{tab:Identified_Lines}. 

\begin{deluxetable*}{llccccc}[ht]
  \tabletypesize{\footnotesize}
  \tablewidth{\textwidth}
  \tablecaption{XCLASS Fitting Results \label{tab:XCLASS}}
  \tablehead{
    \colhead{Species} & 
    \colhead{Formula} & 
    \multicolumn{4}{c}{Column Density [10$^{17}$cm$^{-2}$] 
    }
  }
  \startdata
  &&2014 (88 K)\tablenotemark{a} &2016 (160 K) &2018 (162 K) &2023 (149 K)\\
  \hline
  Methanol & CH$_{3}$OH \tablenotemark{\dag}& 350{\raisebox{0.5ex}{\tiny$\substack{+16 \\ -54}$}}& 130{\raisebox{0.5ex}{\tiny$\substack{+16 \\ -10}$}}& 160{\raisebox{0.5ex}{\tiny$\substack{+7 \\ -5}$}}& 120{\raisebox{0.5ex}{\tiny$\substack{+11 \\ -11}$}} (120{\raisebox{0.5ex}{\tiny$\substack{+10 \\ -32}$}})\tablenotemark{b}\\
 & CH$_{2}$DOH& 6.5{\raisebox{0.5ex}{\tiny$\substack{+0.1 \\ -0.2}$}}& 16{\raisebox{0.5ex}{\tiny$\substack{+1 \\ -0}$}}& 20{\raisebox{0.5ex}{\tiny$\substack{+0 \\ -1}$}}& 15{\raisebox{0.5ex}{\tiny$\substack{+1 \\ -1}$}} (16{\raisebox{0.5ex}{\tiny$\substack{+1 \\ -2}$}})\\
 & CHD$_{2}$OH \tablenotemark{\dag}& 1.8{\raisebox{0.5ex}{\tiny$\substack{+0.2 \\ -3.3}$}}& 4.7{\raisebox{0.5ex}{\tiny$\substack{+0.1 \\ -0.5}$}}& 4.7{\raisebox{0.5ex}{\tiny$\substack{+0.2 \\ -0.1}$}}& 3.2{\raisebox{0.5ex}{\tiny$\substack{+0.2 \\ -0.3}$}} (4.0{\raisebox{0.5ex}{\tiny$\substack{+0.1 \\ -0.2}$}})\\
 & $^{13}$CH$_{3}$OH& 0.40{\raisebox{0.5ex}{\tiny$\substack{+0.01 \\ -0.02}$}}& 3.3{\raisebox{0.5ex}{\tiny$\substack{+0.1 \\ -0.2}$}}& 4.3{\raisebox{0.5ex}{\tiny$\substack{+0.1 \\ -0.1}$}}& 3.5{\raisebox{0.5ex}{\tiny$\substack{+0.1 \\ -0.2}$}} (3.5{\raisebox{0.5ex}{\tiny$\substack{+0.0 \\ -0.4}$}})\\
Ethanol & C$_{2}$H$_{5}$OH& 0.78{\raisebox{0.5ex}{\tiny$\substack{+0.01 \\ -0.02}$}}& 1.9{\raisebox{0.5ex}{\tiny$\substack{+0.0 \\ -0.1}$}}& 2.5{\raisebox{0.5ex}{\tiny$\substack{+0.0 \\ -0.0}$}}& 1.8{\raisebox{0.5ex}{\tiny$\substack{+0.0 \\ -0.2}$}} (2.3{\raisebox{0.5ex}{\tiny$\substack{+0.1 \\ -0.1}$}})\\
Methyl Formate & CH$_{3}$OCHO& 1.2{\raisebox{0.5ex}{\tiny$\substack{+0.2 \\ -0.3}$}}& 1.4{\raisebox{0.5ex}{\tiny$\substack{+0.1 \\ -0.1}$}}& 1.9{\raisebox{0.5ex}{\tiny$\substack{+0.1 \\ -0.1}$}}& 1.7{\raisebox{0.5ex}{\tiny$\substack{+0.1 \\ -0.1}$}} (1.8{\raisebox{0.5ex}{\tiny$\substack{+0.1 \\ -0.1}$}})\\
Dimethyl Ether & CH$_{3}$OCH$_{3}$ \tablenotemark{\dag}& 0.56{\raisebox{0.5ex}{\tiny$\substack{+0.02 \\ -0.02}$}}& 3.1{\raisebox{0.5ex}{\tiny$\substack{+0.2 \\ -0.2}$}}& 5.6{\raisebox{0.5ex}{\tiny$\substack{+0.1 \\ -0.1}$}}& 2.1{\raisebox{0.5ex}{\tiny$\substack{+0.1 \\ -0.2}$}} (1.9{\raisebox{0.5ex}{\tiny$\substack{+0.1 \\ -0.3}$}})\\
Thioformaldehyde & H2CS \tablenotemark{\dag} & 0.13{\raisebox{0.5ex}{\tiny$\substack{+0.01 \\ -0.02}$}}& 0.45{\raisebox{0.5ex}{\tiny$\substack{+0.01 \\ -0.01}$}}& 0.61{\raisebox{0.5ex}{\tiny$\substack{+0.01 \\ -0.01}$}}& 0.51{\raisebox{0.5ex}{\tiny$\substack{+0.01 \\ -0.01}$}} (0.52{\raisebox{0.5ex}{\tiny$\substack{+0.03 \\ -0.01}$}})\\
Sulfur Dioxide & SO$_{2}$ \tablenotemark{\dag}& 0.26{\raisebox{0.5ex}{\tiny$\substack{+0.03 \\ -0.01}$}}& 0.98{\raisebox{0.5ex}{\tiny$\substack{+0.01 \\ -0.01}$}}& 1.6{\raisebox{0.5ex}{\tiny$\substack{+0.1 \\ -0.2}$}}& 1.0{\raisebox{0.5ex}{\tiny$\substack{+0.0 \\ -0.0}$}} (0.90{\raisebox{0.5ex}{\tiny$\substack{+0.01 \\ -0.03}$}})\\
Formamide & HC(O)NH$_{2}$& 0.026{\raisebox{0.5ex}{\tiny$\substack{+0.002 \\ -0.007}$}}& 0.16{\raisebox{0.5ex}{\tiny$\substack{+0.00 \\ -0.00}$}}& 0.20{\raisebox{0.5ex}{\tiny$\substack{+0.01 \\ -0.01}$}}& 0.11{\raisebox{0.5ex}{\tiny$\substack{+0.01 \\ -0.01}$}} (0.10{\raisebox{0.5ex}{\tiny$\substack{+0.02 \\ -0.01}$}})\\
 & DC(O)NH$_{2}$& -& 0.037{\raisebox{0.5ex}{\tiny$\substack{+0.002 \\ -0.003}$}}& 0.053{\raisebox{0.5ex}{\tiny$\substack{+0.001 \\ -0.003}$}}& 0.045{\raisebox{0.5ex}{\tiny$\substack{+0.002 \\ -0.002}$}} (0.045{\raisebox{0.5ex}{\tiny$\substack{+0.014 \\ -0.003}$}})\\
Cyanamide & NH$_{2}$CN \tablenotemark{\dag} & 1.3{\raisebox{0.5ex}{\tiny$\substack{+0.2 \\ -0.6}$}}& 4.3{\raisebox{0.5ex}{\tiny$\substack{+0.1 \\ -0.2}$}}& 5.4{\raisebox{0.5ex}{\tiny$\substack{+0.1 \\ -0.1}$}}& 4.5{\raisebox{0.5ex}{\tiny$\substack{+0.1 \\ -0.1}$}} (3.5{\raisebox{0.5ex}{\tiny$\substack{+0.3 \\ -0.2}$}})\\
Cyanoacetylene & DCCCN \tablenotemark{\dag}& 0.010{\raisebox{0.5ex}{\tiny$\substack{+0.000 \\ -0.001}$}}& 0.014{\raisebox{0.5ex}{\tiny$\substack{+0.015 \\ -0.002}$}}& 0.015{\raisebox{0.5ex}{\tiny$\substack{+0.007 \\ -0.001}$}}& 0.013{\raisebox{0.5ex}{\tiny$\substack{+0.011 \\ -0.005}$}} (0.013{\raisebox{0.5ex}{\tiny$\substack{+0.011 \\ -0.005}$}})\\
  \enddata
  \tablenotetext{\dag}{Species containing only a single transition which contains a distinguishable peak over 3$\sigma$ so that it can be fitted in the spectral range. The CH$_3$OH column density derived by fitting only one line does not seem very reliable, in particular, for the first epoch when compared with the fitting result of $^{13}$CH$_3$OH.}
  \tablenotetext{a}The values inside the parentheses in this row are the excitation temperatures adopted in the XCLASS fitting. We assumed the same source size  ($FWHM=0.12\arcsec$) for all epochs, as described in the main text. 
  \tablenotetext{b}The values inside the parentheses in this column, from this row and below, are the results when all spectra in the full spectral coverage of the fourth epoch (2023) were adopted in the XCLASS fitting.
\end{deluxetable*}
  \label{table:Ncom}

\begin{longtable*}[ht]{lccccc}
  \caption{Identified Lines from the 2023 Spectra} 
  \tablewidth{\textwidth}
  \label{tab:Identified_Lines} \\
  \hline
  \colhead{Formula} & \colhead{Frequency [GHz]} & \colhead{Transition} & \colhead{Einstein-A [log$_{10}A$]} & \colhead{$E_u$ [K]} & \colhead{g$_u$} \\
  \hline
  \endfirsthead

  \multicolumn{6}{c}{{\tablename\ \thetable{} -- continued from previous page}} \\
  \hline
  \colhead{Formula} & \colhead{Frequency [GHz]} & \colhead{Transition} & \colhead{Einstein-A [log$_{10}A$]} & \colhead{$E_u$ [K]} & \colhead{g$_u$} \\
  \hline
  \endhead

  \hline \multicolumn{6}{r}{{Continued on next page}} \\ \hline
  \endfoot

  \hline
  \endlastfoot

  \multicolumn{6}{c}{Methanol} \\
    \hline
    CH$_{3}$OH & 342.729796 & 13(1,12) - 13(0,13)A & 227.47312 & -3.37356 & 108 \\
     & 356.626667 & 23(4,20) - 22(5,18), E & 727.82556 & -4.09468 & 188 \\
     & 356.874537 & 18(8,11) - 19(7,12)E & 717.88089 & -4.31299 & 148 \\
    CH$_{2}$DOH & 342.435476 & 20(2,19) - 20(1,20), o1 & 484.82439 & -3.87291 & 41 \\
     & 342.521696 & 11(6,6) - 12(5,7), e0 & 281.16574 & -4.72978 & 23 \\
     & 342.648908 & 3(2,2) - 2(1,1), e1 & 39.43273 & -4.37753 & 7 \\
     & 342.935645 & 17(2,16) - 17(1,17), e0 & 342.83026 & -3.77664 & 35 \\
     & 345.025252 & 24(4,20) - 24(3,22), e1 & 714.14547 & -3.87467 & 49 \\
     & 345.210673 & 23(4,20) - 23(3,20), e1 & 662.82279 & -3.89104 & 47 \\
     & 345.398904 & 16(2,14) - 15(3,13), e0 & 310.21523 & -4.23438 & 33 \\
     & 345.701510 & 23(4,19) - 23(3,21), e1 & 662.81685 & -3.87425 & 47 \\
     & 345.718718 & 3(2,1) - 2(1,2), e1 & 39.43488 & -4.37316 & 7 \\
     & 354.524281 & 17(3,14) - 17(2,15), e1 & 372.61840 & -4.54869 & 35 \\
     & 354.691156 & 16(2,14) - 15(3,13), e1 & 318.93421 & -4.82735 & 33 \\
     & 356.303575 & 10(2,9) - 10(0,10), o1 & 153.25622 & -4.84790 & 21 \\
     & 356.736665 & 8(7,2) - 7(7,1), o1 & 283.98369 & -4.39296 & 17 \\
     & 356.736665 & 8(7,1) - 7(7,0), o1 & 283.98369 & -4.39296 & 17 \\
     & 356.796142 & 8(6,3) - 7(6,2), o1 & 234.12152 & -4.06757 & 17 \\
     & 356.796142 & 8(6,2) - 7(6,1), o1 & 234.12152 & -4.06757 & 17 \\
     & 356.899665 & 8(2,7) - 7(2,6), e1 & 103.67562 & -3.74509 & 17 \\
     & 356.905068 & 8(5,4) - 7(5,3), o1 & 192.93386 & -3.93737 & 17 \\
     & 356.905068 & 8(5,3) - 7(5,2), o1 & 192.93386 & -3.93737 & 17 \\
     & 356.914710 & 8(2,7) - 7(2,6), o1 & 112.57969 & -3.72878 & 17 \\
     & 356.932418 & 8(4,4) - 7(4,3), e1 & 149.20816 & -3.85223 & 17 \\
     & 356.932437 & 8(4,5) - 7(4,4), e1 & 149.20816 & -3.85223 & 17 \\
     & 356.973848 & 8(4,4) - 7(4,3), o1 & 159.17863 & -3.85225 & 17 \\
     & 356.973870 & 8(4,5) - 7(4,4), o1 & 159.17863 & -3.85225 & 17 \\
     & 356.981575 & 8(0,8) - 7(0,7), o1 & 95.45271 & -3.75309 & 17 \\
     & 356.989824 & 8(3,6) - 7(3,5), e1 & 122.01641 & -3.78413 & 17 \\
     & 356.994178 & 8(3,5) - 7(3,4), e1 & 122.01677 & -3.78412 & 17 \\
     & 357.006964 & 8(2,7) - 7(2,6), e0 & 93.30957 & -3.85897 & 17 \\
     & 357.020783 & 8(3,6) - 7(3,5), o1 & 132.41730 & -3.78354 & 17 \\
     & 357.048396 & 8(6,3) - 7(6,2), e1 & 228.95500 & -4.14455 & 17 \\
    CHD$_{2}$OH & 343.057196 & 3(2)-3(1)+ & 31.83515 & -3.85643 & 7 \\
     & 343.151347 & 16(2)+ - 16(2)- & 302.90566 & -4.12444 & 33 \\
     & 345.445553 & 15(1)- - 15(1)+ & 261.23105 & -3.66430 & 31 \\
     & 355.100973 & 5(2)+ - 5(1)- & 49.82160 & -3.79463 & 11 \\
    $^{13}$CH$_{3}$OH & 345.132599 & 4(0,4) - 3(-1,3) & 35.76015 & -4.08423 & 9 \\
     & 354.445952 & 4(1,3) - 3(0,3) & 43.71183 & -3.89468 & 9 \\
     & 356.873814 & 13(2,12) - 12(3,9) & 243.83835 & -4.10131 & 27 \\
  \hline 
  \multicolumn{6}{c}{Ethanol} \\
    \hline
    C$_{2}$H$_{5}$OH & 345.173949 & 7(7,1) - 6(6,1), g+ & 139.90163 & -3.59865 & 15 \\
     & 345.173949 & 7(7,0) - 6(6,0), g+ & 139.90163 & -3.59865 & 15 \\
     & 345.229285 & 21(1,21) - 20(1,20), g+ & 241.55353 & -3.43030 & 43 \\
     & 345.295355 & 21(1,21) - 20(1,20), g- & 246.22196 & -3.43536 & 43 \\
     & 345.333442 & 21(0,21) - 20(0,20), g+ & 241.54097 & -3.42989 & 43 \\
     & 345.408165 & 21(0,21) - 20(0,20), g- & 246.20853 & -3.43484 & 43 \\
     & 345.648571 & 20(3,18) - 19(3,17), g+ & 242.48944 & -3.42195 & 41 \\
     & 345.656622 & 20(3,18) - 19(3,17), g- & 247.21350 & -3.45470 & 41 \\
     & 354.363234 & 20(3,17) - 19(2,17), g+ & 244.98529 & -3.31183 & 41 \\
     & 354.757875 & 20(3,17) - 19(3,16), g- & 249.36317 & -3.43762 & 41 \\
     & 357.067425 & 10(4,7) - 9(3,6), anti & 66.31120 & -3.64071 & 21 \\
     & 357.119887 & 21(1,20) - 20(1,19), g+ & 250.89431 & -3.38089 & 43 \\
    \multicolumn{6}{c}{Methyl Formate} \\
    \hline
    CH$_{3}$OCHO & 343.148047 & 31(2,30) - 30(2,29)E & 273.44228 & -3.21580 & 126 \\
     & 345.069059 & 28(14,15) - 27(14,14)A & 369.64307 & -3.32711 & 114 \\
     & 345.465345 & 16(13,4) - 16(12,5)E & 192.32292 & -4.59986 & 66 \\
     & 345.466962 & 28(13,16) - 27(13,15)A & 351.85611 & -3.30627 & 114 \\
     & 354.607764 & 33(1,33) -32(1,32)E & 293.11551 & -3.14319 & 134 \\
     & 354.607765 & 33(0,33) - 32(0,32), E & 293.11551 & -3.14319 & 134 \\
    \hline
    \multicolumn{6}{c}{Dimethyl Ether} \\
    \hline
    CH$_{3}$OCH$_{3}$ & 342.608177 & 19(0,19)-18(1,18)AA & 167.14029 & -3.28159 & 195 \\
     & 356.443701 & 35(7,28) - 35(6,29)AA & 643.55343 & -3.39929 & 355 \\
     & 356.575254 & 8(4,5) - 7(3,4)AE & 55.26662 & -3.67917 & 51 \\
     & 356.582852 & 8(4,5) - 7(3,4)AA & 55.26669 & -3.67911 & 85 \\
     & 356.724446 & 8(4,4)- 7(3,5)AA & 55.26673 & -3.67880 & 51 \\
    \hline
    \multicolumn{6}{c}{Thioformaldehyde} \\
    \hline
    H$_{2}$CS & 342.946424 & 10(0,10) - 9(0,9) & 90.59358 & -3.21609 & 21 \\
     & 343.203239 & 10(5,5) - 9(5,4) & 419.18398 & -3.34004 & 63 \\
     & 343.203239 & 10(5,6) - 9(5,5) & 419.18398 & -3.34004 & 63 \\
     & 343.309830 & 10(4,7) - 9(4,6) & 301.08004 & -3.29045 & 21 \\
     & 343.309830 & 10(4,6) - 9(4,5) & 301.08004 & -3.29045 & 21 \\
     & 343.322082 & 10(2,9) - 9(2,8) & 143.31041 & -3.23242 & 21 \\
    \hline
    \multicolumn{6}{c}{Sulfur Dioxide} \\
    \hline
    SO$_{2}$ & 342.761625 & 34(3,31) - 34(2,32) & 581.91876 & -3.46261 & 69 \\
     & 345.338538 & 13(2,12) - 12(1,11) & 92.98367 & -3.62327 & 27 \\
     & 355.045517 & 12(4, 8) - 12(3, 9) & 110.99905 & -3.46907 & 25 \\
     & 356.755190 & 10(4,6) - 10(3,7) & 89.83365 & -3.48406 & 21 \\
     & 357.165390 & 13(4,10) - 13(3,11) & 122.96459 & -3.45446 & 27 \\
    \hline
    \multicolumn{6}{c}{Formamide} \\
    \hline
    HC(O)NH$_{2}$ & 343.083117 & 16(3,13) - 15(3,12) & 165.99525 & -2.54136 & 33 \\
     & 343.196986 & 17(1,17) - 16(1,16) & 151.99629 & -2.52664 & 35 \\
     & 345.181258 & 17(0,17) - 16(0,16) & 151.58895 & -2.51879 & 35 \\
     & 345.325391 & 16(1,15) - 15(1,14) & 145.15586 & -2.52018 & 33 \\
     & 356.713759 & 17(2,16) - 16(2,15) & 166.78985 & -2.48082 & 35 \\
    HC(O)NH$_{2}$, v$_{t}$=1 & 342.735819 & 16(3,13) - 15(3,12) & 581.24411 & -2.54275 & 33 \\
     & 343.224648 & 17(1,17) - 16(1,16) & 567.73995 & -2.52645 & 35 \\
     & 344.964708 & 16(1,15) - 15(1,14) & 560.72867 & -2.52151 & 33 \\
     & 345.182139 & 17(0,17) - 16(0,16) & 567.33737 & -2.51874 & 35 \\
     & 356.502153 & 17(2,16) - 16(2,15) & 582.28981 & -2.48157 & 35 \\
    DC(O)NH$_{2}$ & 345.511808 & 17(2,16) - 16(2,15) & 159.40228 & -2.52324 & 105 \\
     & 354.416031 & 17(8,10) - 16(8,9) & 289.63063 & -2.59138 & 105 \\
     & 354.416032 & 17(8,9) - 16(8,8) & 289.63063 & -2.59138 & 105 \\
     & 354.549286 & 17(3,15) - 16(3,14) & 172.50787 & -2.49637 & 105 \\
     & 354.661332 & 17(7,10) - 16(7,9) & 257.67084 & -2.56245 & 105 \\
     & 355.059342 & 17(6,11) - 16(6,10) & 230.00266 & -2.53804 & 105 \\
     & 356.382719 & 17(4,14) - 16(4,13) & 187.64341 & -2.50029 & 105 \\
    \hline
    \multicolumn{6}{c}{Cyanamide} \\
    \hline
    NH$_{2}$CN & 345.285952 & 10(1,10) - 11(2,10), v=1-0 & 137.85691 & -4.13466 & 21 \\
    \multicolumn{6}{c}{Cyanoacetylene} \\
    \hline
    HC$_{3}$N & 345.609010 & J=38-37 & 323.49156 & -2.48124 & 77 \\
     & 354.697463 & J=39-38 & 340.51429 & -2.44728 & 79 \\
    DCCCN & 354.478929 & J=42-41 & 365.82987 & -2.44555 & 85 \\
    \hline
    \multicolumn{6}{c}{Acetaldehyde} \\
    \hline
    CH$_{3}$CHO & 354.812939 & 18(2,16) - 17(2,15)E & 169.65483 & -2.80104 & 74 \\
     & 354.844371 & 18(2,16) - 17(2,15)A & 169.64051 & -2.80112 & 74 \\
    CH$_{3}$CHO, v$_{t}$=1 & 354.457665 & 18(2,16) - 17(2,15)A & 374.96403 & -2.80035 & 74 \\
     & 354.525390 & 19(0,19)-18(0,18)E & 376.56819 & -2.80091 & 78 \\
    \hline
    \multicolumn{6}{c}{Formaldehyde} \\
    \hline
    D$_{2}$CO & 342.522128 & 6(0,6) - 5(0,5) & 58.11953 & -2.93313 & 26 \\
    H$_{2}^{13}$CO & 343.325713 & 5(1,5) - 4(1,4) & 61.28076 & -2.95171 & 33 \\
     & 354.898595 & 5(2,4) - 4(2,3) & 98.41454 & -2.96644 & 11 \\
     & 355.028955 & 5(4,1) - 4(4,0) & 240.14118 & -3.33392 & 11 \\
     & 355.028955 & 5(4,2) - 4(4,1) & 240.14118 & -3.33392 & 11 \\
    \hline
    \multicolumn{6}{c}{Formic Acid} \\
    \hline
    t-HCOOH & 342.521194 & 16(1,16) - 15(1,15) & 143.59489 & -3.34109 & 33 \\
     & 345.030561 & 16(0,16) - 15(0,15) & 143.05621 & -3.33118 & 33 \\
    \hline
    \multicolumn{6}{c}{Ketene} \\
    \hline
    H$_{2}$CCO & 343.088615 & 17(5,12) - 16(5,11) & 473.86060 & -3.37492 & 105 \\
     & 343.088615 & 17(5,13) - 16(5,12) & 473.86060 & -3.37492 & 105 \\
     & 343.172572 & 17(0,17) - 16(0,16) & 148.30864 & -3.33525 & 35 \\
     & 343.250411 & 17(4,14) - 16(4,13) & 356.84905 & -3.35968 & 35 \\
     & 343.250411 & 17(4,13) - 16(4,12) & 356.84905 & -3.35968 & 35 \\
    \hline
\end{longtable*}
\onecolumngrid
\vspace{-5 mm}

\end{document}